\documentclass{elsart}
\usepackage{amsmath}
\usepackage{amssymb}
\usepackage{epsfig}
\begin{document}
\begin{frontmatter}
\title{Beyond mean-field description\\ 
       of the low-lying spectrum of $^{16}$O}
\author{M. Bender and P.-H. Heenen}
\address{Service de Physique Nucl\'{e}aire Th\'{e}orique 
         et de Physique Math\'{e}matique\\
         Universit{\'e} Libre de Bruxelles 
         -- C.P.\ 229, B-1050 Brussels, Belgium}
\date{September 2 2002}
\maketitle
\begin{abstract}
Starting from constrained Skyrme-mean-field calculations, the low-energy 
excitation spectrum of $^{16}$O is calculated by configuration mixing 
of particle-number and angular-momentum projected mean-field states
in the framework of the Generator Coordinate Method. Without any 
adjustable parameters, this approach gives a very good description 
of those states and their transition moments that can be described 
with our restriction to axially and reflection-symmetric shapes.
The structure of low-lying $0^+$ states is analyzed in terms
of self-consistent 0p-0h, 2p-2h, and 4p-4h Hartree-Fock states. 
 \end{abstract}
\end{frontmatter}
%
%
\section{Introduction}
The doubly-magic oxygen isotope $^{16}$O has always attracted much 
interest in nuclear structure physics. Its first excited $0^+$ state 
has been interpreted by Morinaga \cite{Mor56} 
as an evidence for shape coexistence. Multi-nucleon transfer
reactions have lead to the conclusion  that the presence
of a state corresponding to a deformed shape is due to excitations 
of both proton and neutron pairs across the $N$ and \mbox{$Z=8$}
closed shells. Several other excited states at low excitation energies 
can be interpreted as due to multi-particle-hole excitations (see the 
discussion in the Wood \emph{et al.\/} review paper \cite{Woo92}). 
The first excited $0^+$ state is the head of a rotational band.

Shell model and mean-field calculations were already
performed in the sixties \cite{Bas65,Bro66,Kri69} with a
description of this $0^+$ state  based on a deformed 4p-4h 
configuration. The cluster model \cite{Ber71}, 
using an empirical $\alpha$-$\alpha$ interaction, qualitatively explains
the deformed structure of the excited states. To go beyond a 
qualitative understanding of the $^{16}$O spectrum appeared  however
very quickly as a difficult challenge. 
Mean-field calculations with more modern effective interactions 
\cite{Gir83,Auv84,Zhe88} clearly showed that correlations beyond 
the Hartree-Fock (HF) approach have to be included to describe 
successfully the first excited $0^+$ state. Two kinds of correlations 
were mainly invoked: pairing correlations,
and restoration of rotational symmetry which both leave the
spherical configuration unchanged and lower the energy of
the deformed configurations. 

More recently, shell model calculations mixing (0+2+4)$\hbar\omega$
excitations \cite{Hax90,Wab92} have confirmed the schematic model
of Brown and Green \cite{Bro66}, and showed that the large quadrupole
moment of the deformed configuration appears naturally in a full shell
model diagonalization. It has also been possible in such
calculations to test how well
$^{16}$O can be considered as an inert core as it is usually done in
shell model calculations. The 0p-0h component of the ground-state 
wave function was found to have a weight of the order of 40 to 50$\%$ 
only, with nearly equal importance of the 2p-2h components.

The question of a correct description of the $^{16}$O spectrum
starting from a mean-field approach is thus still an open question.
This nucleus is an anchor point where many approaches to the nuclear
many-body problem can be compared. Indeed, besides the already 
mentioned mean-field and shell model calculations using effective 
interactions, there are also calculations based on so-called ``realistic'' 
nucleon-nucleon interactions (introduced as effective potentials 
to describe the phase shifts in nucleon-nucleon scattering), 
either already performed for $^{16}$O \cite{Fab00}, or to be expected 
for the near future \cite{Nav00}.

In this paper, we present an application of the method introduced by 
Valor \emph{et al.\/} \cite{Val00}, which performs a configuration 
mixing of projected self-consistent mean-field states, to calculate
the low-energy spectrum of $^{16}$O, and to analyze its structure 
in terms of $n$p-$n$h Hartree-Fock states. 
%
%
\section{The Method}
The starting point of the method is a set of BCS states
$|q \rangle$ generated by constrained Skyrme-mean-field calculations.
We have used the Skyrme parametrization SLy4 \cite{Cha98} together with 
a like-particle \mbox{$T=1$} density-dependent zero-range pairing 
interaction \cite{Ter96}. The pairing strength adjusted in \cite{Rig99} 
has been reduced from $V = - 1250$ MeV fm$^{3}$ 
to $V = - 1000$ MeV fm$^{3}$ for both protons and  neutrons, as in 
\cite{Val00}. The pairing active space is limited by a soft 
cutoff at 5 MeV above and below the Fermi energy \cite{Bon85}.
This combination of effective interactions has been proven to be very 
successful in the description of a large number of experimental data 
all over the chart of nuclei. 

The set of mean-field wave functions $|q \rangle$ is generated by 
calculations with a constraint on the axial quadrupole moment $Q_0$. 
To avoid a collapse of pairing correlations along the constraining path, 
the approximate variation-after-projection Lipkin-Nogami (LN) method is 
used. The Lipkin-Nogami prescription to evaluate an energy correction 
due to particle number projection is known to have deficiencies when
pairing correlations are weak \cite{Mag93}. We do not make use of this
correction, however, but introduce the LN method solely to ensure that 
pairing correlations, although weak, are present even in the spherical 
configuration. With that, the LN method provides BCS wave functions which 
are a fair starting point for an exact projection on particle number. 
Since these projected BCS wave functions are used further as 
basis functions for a configuration mixing, the 
deficiencies of the LN method should not affect the results.

The most important symmetries broken by the mean-field approach are 
restored after variation by standard projection techniques \cite{Val00}. 
For each value of the quadrupole moment, the wave functions are 
projected simultaneously on angular momentum 
and on neutron and proton particle numbers, decomposing a given 
intrinsic BCS state $|q \rangle$ into wave functions $| J q \rangle$ 
corresponding to several values of the angular momentum. 
For each value of $J$ separately, these sets of non-orthogonal states 
$| J q \rangle$ are then mixed as a function of the quadrupole moment
to give the final wave functions $|J k \rangle$, which correspond
to collective states of the nucleus. A discrete number of values of 
the quadrupole moment are considered, but in such a way that the results
do not depend on the discretization and are equivalent to a mixing
on a continuous variable as in the generator coordinate method (GCM)
\cite{Hil53,Bon90}. Among the configuration-mixed states $| J k \rangle$ 
so obtained, the physically interesting ones are the ground 
state ($J=0$, $k=0$) and the few excited states which can be described
by an axial quadrupole collective mode. States corresponding to modes 
that require a breaking of reflection symmetry (like $\alpha$-$^{12}$C
configurations or the tetrahedron-like 4-$\alpha$ configuration) 
are not included in our description of $^{16}$O. 
%
%
\section{Projection}
%
%
\begin{figure}[t!]
\centerline{\epsfig{file=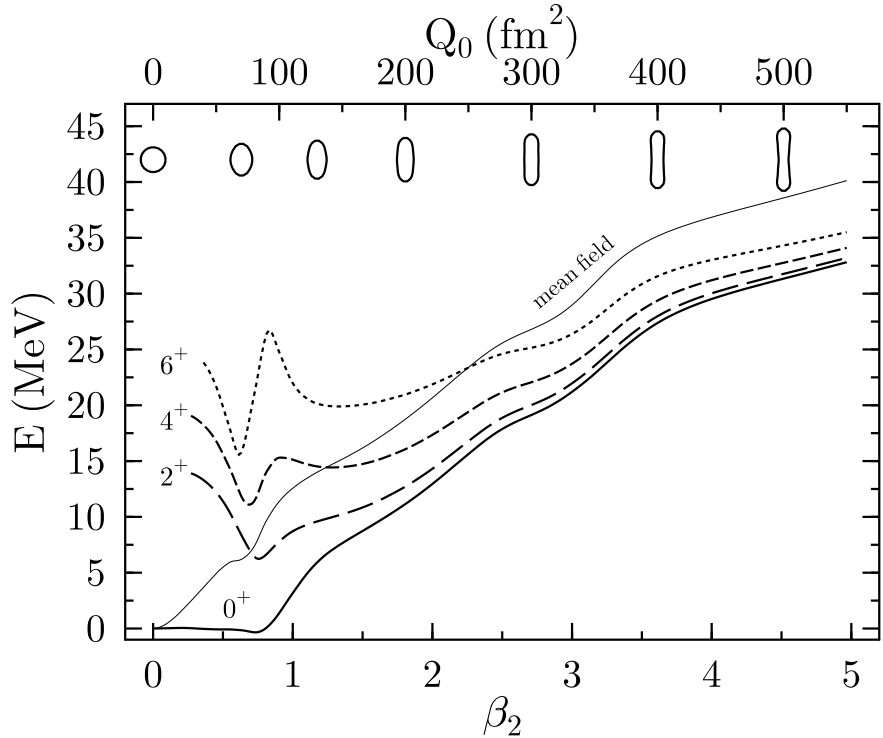,width=12cm}}
\caption{\label{fig:e}
Particle-number projected (``mean field'') and particle-number 
and angular-momentum projected potential energy curves for $^{16}$O. 
All curves are drawn versus the (intrinsic) deformation of the BCS 
state which is projected. The upper axis gives the mass quadrupole
moment $Q_0 = \sqrt{16 \pi/5} \, \langle r^2 Y_{20} \rangle$, 
the lower the dimensionless deformation parameter 
$\beta_2 = 4 \pi \langle r^2 Y_{20} \rangle /(3 R^2 A)$ 
with $R = 1.2 \, A^{1/3}$ fm. 
The various shapes along the paths are indicated by the contours
of the total density at $\rho_0 = 0.07 \, {\rm fm}^{-3}$.
}
\end{figure}
%
%
%
%
\begin{figure}[t!]
\centerline{\epsfig{file=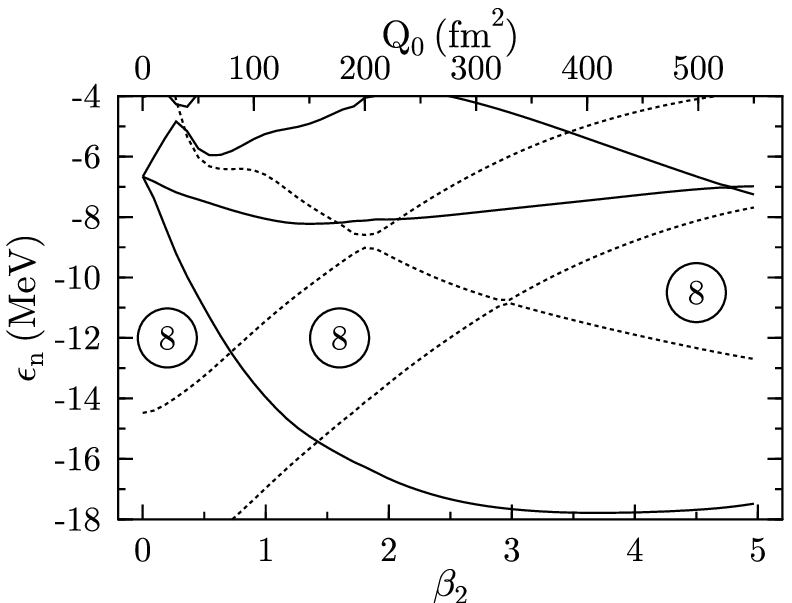,width=10cm}}
\caption{\label{fig:s}
Single-particle spectra for neutrons in $^{16}$O.
Solid (dotted) lines represent levels with positive (negative) 
parity.
}
\end{figure}
%
%
The energy curves are shown in figure \ref{fig:e}. The mean-field curve, 
projected on particle numbers \mbox{$N=Z=8$} only, shows, as expected, a 
deep minimum for the spherical configuration and no secondary minima. 
Two inflection points, however, are present at large deformations 
(80 and 150 fm$^2$). Looking at the single-particle level scheme 
plotted for neutrons 
in figure \ref{fig:s}, one sees that these points correspond 
to level crossings. At a $\beta_2$ value around 0.8 (or a mass 
quadrupole moment equal to 80 fm$^2$), a $d_{5/2}$ level crosses the 
$p_{1/2}$ level and is located below the Fermi level for
larger quadrupole moments. Since due to time-reversal invariance,
the levels are doubly degenerate  and since 
neutrons and protons have a very similar behavior in this $N=Z$ nucleus, 
the mean-field wave functions for $\beta_2$ 
values larger than 0.8 correspond to a deformed 4p-4h configuration. 
Owing to pairing correlations, however, all the levels around the 
Fermi energy have sizable occupation probabilities. In the same way, 
for a $\beta_2$ value around 1.8, a crossing occurs between a state 
coming down from the $pf$ shell and a $p$ state. Note that for the 
protons the level crossings are at slightly smaller deformation than 
for the neutrons, a consequence of the Coulomb interaction. For 
extremely deformed shapes, above \mbox{$\beta_2=4.0$}, the mean-field
configuration corresponds to a deformed 8p-8h configuration
which is associated to a chain of four $\alpha$ particles. 

The energy curves projected on angular momentum, from \mbox{$J=0$} to 
\mbox{$J=6$}, are also shown on figure \ref{fig:e}. While the spherical 
mean-field state has a \mbox{$J=0$} component only, the deformed 
configurations can be projected on several angular momenta. Their \mbox{$J=0$} 
components have always energies lower than the mean-field energy. 
Since for each quadrupole moment the mean-field energy is the weighted 
sum of the projected energies, figure \ref{fig:e} shows also that 
components with \mbox{$J > 6$} are dominant beyond 
\mbox{$\beta_2 \approx 2.5$}. 

The \mbox{$J=0$} curve is extremely flat up to the first single-particle 
level crossing. Beyond this point, the gain in energy due to angular 
momentum projection is approximately constant and of the order of 7 MeV. 
For angular momenta equal to 2 to 6, a well-developed minimum is obtained 
for mean-field states with an axial quadrupole moment around 75 fm$^2$
which correspond to the deformation at which single-particle level
crossing occurs. The shape of the mean-field density is also changing 
abruptly at this point, as can be seen on figure \ref{fig:e}
and is reflected in the hexadecapole deformation. For a quadrupole
moment value around 100 fm$^2$, the $\beta_4$ parameter, 
as defined in \cite{Cwi96}, changes sign.
For still more deformed intrinsic wave functions, the projected 
energies vary very slowly, with soft minima obtained for $\beta_2$ 
values around 1.5.
%
%
\section{Configuration Mixing}
\begin{table}[t!]
\centerline{
\begin{tabular}{ccccccc} 
\noalign{\smallskip}\hline\noalign{\smallskip}
state   & $E_{\rm expt}$
        & $E_{\rm calc}$
        & $r_{\rm rms}$
        & $Q_s$
        & $Q_0(s)$
        & $\beta_2(s)$ \\
        & (MeV)
        & (MeV)
        & (fm)
        & (e fm$^2$)
        & (e fm$^2$)
        &             \\ 
\noalign{\smallskip}\hline\noalign{\smallskip}
$0^+_1$ &  0.00 & 0.00  & 2.75 &         &      &      \\
$0^+_2$ &  6.05 & 6.03  & 2.90 &         &      &      \\
$2^+_1$ &  6.92 & 7.09  & 2.91 & $-11.7$ & 41.0 & 0.74 \\
$0^+_3$ & 12.05 & 12.45 & 3.32 &         &      &      \\
$2^+_2$ & 13.02 & 13.18 & 3.22 & $-22.9$ & 80.2 & 1.4  \\
$0^+_4$ &  --   & 20.73 & 3.85 &         &      &      \\
$2^+_3$ &  --   & 23.06 & 3.82 & $-44.4$ & 155  & 2.8  \\
\noalign{\smallskip}\hline\noalign{\smallskip}
\end{tabular}}
\caption{\label{tab:ex1}
Experimental ($E_{\rm expt}$) and calculated ($E_{\rm calc}$)
excitation energies for the low-lying $0^+$ and $2^+$ states in $^{16}$O.
Only states that can be described by our calculation are indicated.  
Also given are the calculated proton rms radii $r_{\rm rms}$, 
spectroscopic quadrupole moments $Q_s$ (in the laboratory frame), 
proton quadrupole moments $Q_0(s)$ and  $\beta_2(s)$ deformation 
parameters in the intrinsic frame. Experimental excitation energies 
are taken from \protect\cite{Til93aE}.
}
\end{table}
%
%
\begin{figure}[b!]
\centerline{\epsfig{file=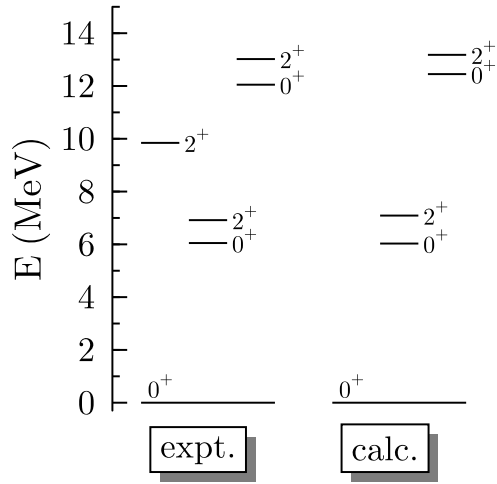,width=8cm}}
\caption{\label{fig:spectra}
Comparison between the experimental and
theoretical low-lying $0^+$ and $2^+$ states in $^{16}$O. 
}
\end{figure}
%
%
%
%
\begin{figure}[t!]
\centerline{\epsfig{file=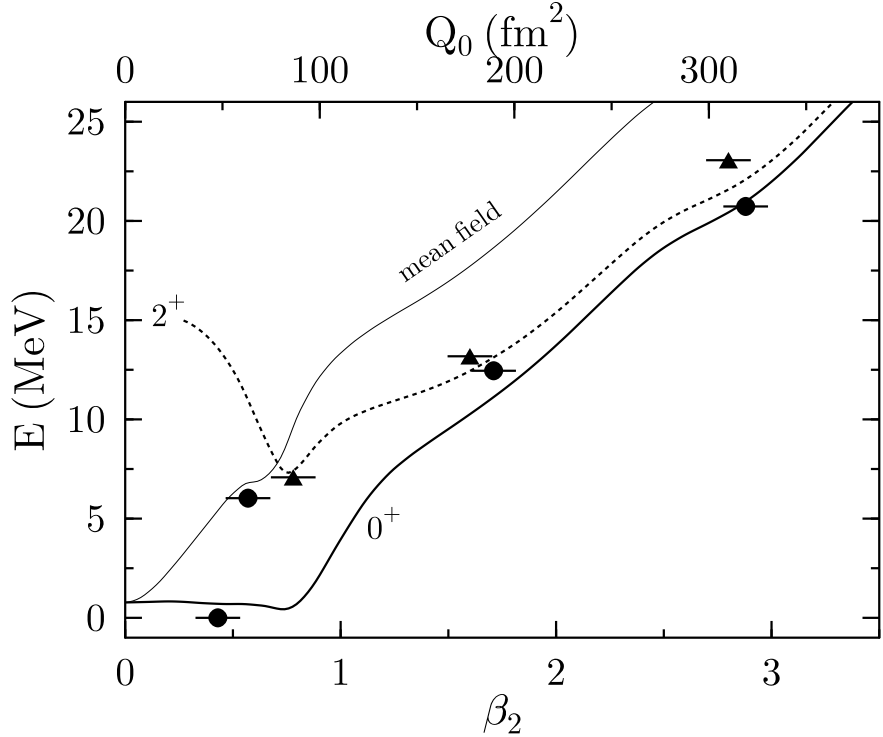,width=12cm}}
\caption{\label{fig:ex1}
Spectrum of the lowest $0^+$ (circles) and $2^+$ (triangles)
states, plotted at a deformation corresponding to the
average deformation of the intrinsic wave functions. These averages
are determined using the weight of each mean-field state in the
collective wave functions.
Also shown are the angular momentum projected energy curves and 
the (particle-number projected only) ``mean-field'' curve (thin black line). 
}
\end{figure}
%
%
%
%
\begin{figure}[t!]
\centerline{\epsfig{file=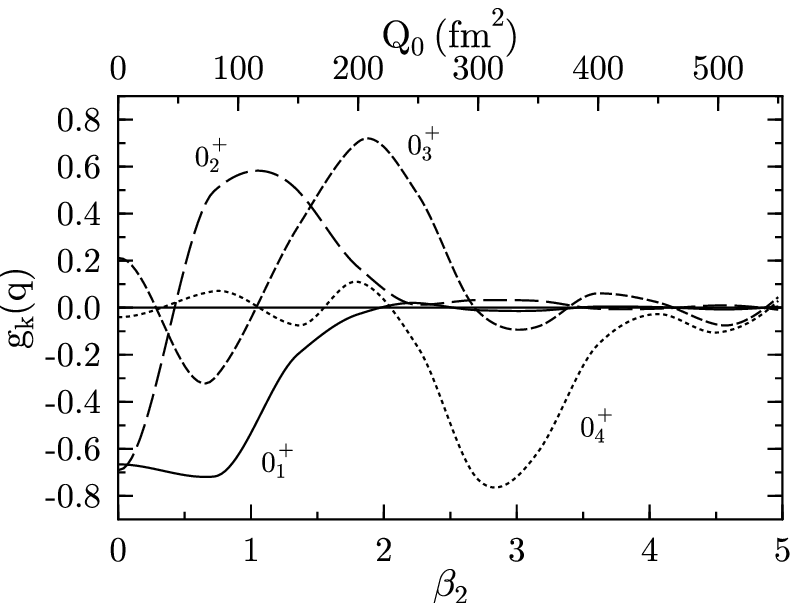,width=12cm}}
\caption{\label{fig:wf0}
GCM wave functions $| J k \rangle = \sum_q g_k(q) | J q \rangle$ of the four lowest $0^+$ states.
}
\end{figure}
%
%
The configuration mixing of mean-field wave functions as performed
in this work has several goals. The particle-number projection
removes unwanted contributions 
coming from states with different particle numbers. The angular
momentum projection separates the contribution from different
angular momenta and generates wave-functions in the laboratory
frame which provide  transition probabilities
and spectroscopic moments without further approximations. Finally, the 
variational configuration mixing with respect to a collective
coordinate, the axial quadrupole moment in this work, removes the 
contributions to the ground state coming from collective vibrations 
and provides the excitation spectrum corresponding to this mode.

Results for the lowest states obtained in the configuration 
mixing calculation are given in table \ref{tab:ex1} and are compared 
with the experimental excitation energies.
The calculated excitation spectrum is compared to the 
experimental data in figure \ref{fig:spectra} and to the mean-field 
energy curve in figure \ref{fig:ex1}. 
Only prolate configurations have been included. 
The projected oblate configurations have an overlap 
close to 1.0 with the spherical configuration which makes the 
configuration mixing calculation numerically unstable. The rms 
radii and deformations given in table \ref{tab:ex1} 
are calculated from the point-proton distribution and do not include 
the corrections that have to be introduced to calculate charge 
radii that can be compared with the experimental data, 
see e.g.\ \protect\cite{RMP}. 

With our method, one obtains automatically the spectroscopic 
quadrupole moments $Q_s$ in the laboratory frame. 
The charge quadrupole moment in the intrinsic 
frame $Q_0(s)$ is then calculated with the relation
\mbox{$Q_0(s) = - Q_s \, (2J+3)/J$} assuming \mbox{$K=0$} 
states \cite{RS80}. The dimensionless deformation parameter 
$\beta_2$ is related to $Q_0(s)$ by
\mbox{$\beta_2(s) = \sqrt{5/(16 \pi)} \times 4 \pi Q_0 (s) /(3 R^2 Z e^2)$}.

The assignment of the calculated $0^+_2$ and $2^+_1$ states to 
experiment in table \ref{tab:ex1} is tentative. 
Note that the restriction of our calculation 
to reflection-symmetric configurations 
does not permit to describe states with negative 
parity or whose intrinsic structure is supposed to be asymmetric.
For instance, the $\alpha$-$^{12}$C configuration assigned to 
the experimental $2^+$ state at 9.8440 MeV \cite{Til93aE}
cannot be represented by our model.
For the other states, the agreement between the experimental 
and theoretical energies is excellent.

The correlations introduced by the configuration mixing decrease the 
energy of the $0^+_1$ state with respect to the HF 
ground state by 2.3 MeV. This gain in energy, although significant, 
is lower than in several of our previous configuration-mixing calculations. 
In $^{208}$Pb, for instance, removing quadrupole vibrations from the
mean-field ground state by configuration mixing without angular
momentum projection brings already 2.0 MeV of extra energy \cite{Hee01}. 
In $^{24}$Mg, restoration of symmetries and the inclusion of the contribution
from quadrupole vibrations decrease the mean-field ground-state 
energy by more than 5 MeV \cite{Val00}.

 The mixing coefficients $g_k(q)$ for the four 
lowest $0^+$ states are plotted in figure \ref{fig:wf0}.
The ground-state wave function is spread over a large range of deformations, 
with nearly equal weight up to a \mbox{$\beta_2 \approx 0.8$}. 
Its rms radius \mbox{$r_{\rm rms} = 2.75$} fm is therefore
significantly larger than the \mbox{$2.68$} fm value
of the spherical HF state. The maximum 
weight for the first excited state is obtained around $\beta_2$ equal 
to 1.0, but with significant components at lower quadrupole moments. 
The third and fourth $0^+$ states are shifted to configurations 
with larger deformations. As can be seen from the mean deformation 
of their mean-field components (see Fig. \ref{fig:ex1}), 
and the shapes displayed in Fig.\ \ref{fig:e}, none of them corresponds 
to a chain of 4 $\alpha$ particles which occurs at even larger deformation 
around \mbox{$\beta_2 \approx 4.0$}.

\begin{table}[t!]
\centerline{
\begin{tabular}[t]{lccccc} 
\noalign{\smallskip}\hline\noalign{\smallskip} 
transition & \multicolumn{2}{c}{$B(E2)\uparrow/(e^2\text{fm}^4)$}
           & $Q_0(t)$
           & $\beta_2 (t)$ \\
\noalign{\smallskip} \cline{2-3}\noalign{\smallskip} 
           &  expt.
           &  calc.
           & ($e$ fm$^2$)
           &           \\ 
\noalign{\smallskip}\hline\noalign{\smallskip}  
$2^+_1 \to 0^+_1$ & $ 42 \pm 1.4$ & $ 38$  & 19.7 & 0.36 \\
$2^+_2 \to 0^+_1$ & $ 21 \pm 7  $ & $4.7$  &  6.9 & 0.12 \\
$2^+_1 \to 0^+_2$ & $370 \pm 4  $ & $241$  & 49.2 & 0.89 \\
$2^+_2 \to 0^+_2$ & $42  \pm 7  $ & $ 74$  & 27.3 & 0.49 \\
$2^+_2 \to 0^+_3$ &     ---       & $590$  & 77.0 & 1.39 \\
\noalign{\smallskip}\hline\noalign{\smallskip} 
\end{tabular}
\begin{tabular}[t]{lccc}
\noalign{\smallskip}\hline\noalign{\smallskip} 
transition & \multicolumn{2}{c}{$M(E0)/$fm$^2$} \\ 
\noalign{\smallskip} \cline{2-3}\noalign{\smallskip} 
           &  expt.           &    calc.        \\ 
\noalign{\smallskip}\hline\noalign{\smallskip}  
$0^+_2 \to 0^+_1$ & $3.55 \pm 0.21$ & 5.735 \\
$0^+_3 \to 0^+_1$ & $4.03 \pm 0.09$ & 0.690 \\
\noalign{\smallskip}\hline\noalign{\smallskip}  
\end{tabular}}
\caption{\label{tab:be}
Reduced transition probabilities $B(E2) \uparrow$, calculated
transition quadrupole moments $Q_0(t) = \sqrt{16 \pi B(E2) \uparrow / 5}$, 
corresponding deformation parameters $\beta_2 (t)$, 
and monopole matrix transition elements
$M(E0) = \langle f | \sum_p r_p^2 | i \rangle$ between low-lying 
states in $^{16}$O. The experimental data are taken from 
\protect\cite{Til93aE}.
}
\end{table}

The $B(E2)$ values for transitions from the $2^+_1$ to the ground
state and to the first excited $0^+$ state are well described
by our model. The error is larger for the transition 
from the $2^+_2$ state to the ground state.  This is not surprising since
important configurations like $\alpha-^{12}$C which are not included
in our model should play a role at the excitation energy of the $2^+_2$ state.
For a similar reason, the monopole transition probabilities are in better
agreement with experiment for the first excited $0^+$ state than for the
second. 

Table \ref{tab:be} lists the experimental value from \cite{Til93aE}.
The values adopted by Raman \emph{et al.\/} \cite{Ram01a} for the 
\mbox{$B(E2) \uparrow = (40.6 \pm 3.8)$ $e^2$ fm$^4$} of the 
$2^+_1 \to 0^+_1$ transition is even in better agreement with our
result. The $Q_0(t)$ of the transitions within a rotational band,
i.e.\ $2^+_1 \to 0^+_2$ and $2^+_2 \to 0^+_3$, are close to the intrinsic 
quadrupole moments $Q_0 (s)$ of the $2^+$ states involved as given in 
table \ref{tab:ex1}.
%
%
\section{Analysis of the Collective States}
Analyses of the shell model wave functions obtained in a (0+2+4)$\hbar \omega$
active space \cite{Hax90,Wab92} have shown that the $^{16}$O ground-state 
wave function is by far not a pure 0$\hbar  \omega$ state and that the 
breaking of the closed shells at $N$ and $Z$ equal to 8 is large. 
The 0$\hbar  \omega$ component of the shell-model wave function has 
a weight of the order of 50 $\%$ only, with components of comparable 
importance in the 2 $\hbar \omega$ subspace. The first excited $0^+$ 
state has an energy very close to the first experimental $0^+$ and 
is composed to 90$\%$ of a 4p-4h configuration.

It is not evident to perform a similar analysis for self-consistent
wave functions. In our model, there is no spherical oscillator basis 
on which the individual wave functions are expanded. 
However, it is important to find a common language with 
shell-model calculations and to interpret a deformed mean-field basis
in terms of a spherical basis. In the framework of self-consistent 
mean-field models, a natural choice for the reference state is 
provided by the self-consistent Hartree-Fock (HF) 0p-0h ground state, 
which is of course spherical for the closed-shell nucleus $^{16}$O.
To build $n$p-$n$h excitations in an HF framework requires some care.
They could be constructed non-self-consistently on top of the 
spherical HF state by just changing the occupation of the 
single-particle levels below and above the Fermi surface. 
However, such excitations are not unique and the resulting wave 
function is, in general, not spherical. A much more natural procedure 
is to consider the deformation also in the $n$p-$n$h HF states and 
to construct them self-consistently by solving the HF equations 
with appropriate occupation numbers of the single-particle states. 
These 4p-4h and 8p-8h states have deformations corresponding to 
the deformed shell closures that can be seen around 
\mbox{$\beta_2 \approx 1.4$} and \mbox{$\beta_2 \approx 4.5$} in 
figure \ref{fig:s}. 
Each fully self-consistent $n$p-$n$h HF state $| \text{$n$p-$n$h} \rangle$ 
defines a basis of deformed single-particle states different from the 
other excitations. As a consequence, the various 
$| \text{$n$p-$n$h} \rangle$ are not orthonormal.
However, since of their very different structure (deformation and $n$p-$n$h
excitations), their overlap is extremely small and can be neglected.

Since they are deformed, the $n$p-$n$h HF states are not eigenstates of
angular momentum and projected $n$p-$n$h states $| J \text{$n$p-$n$h} \rangle$
are constructed by symmetry restoration.

We analyze the $n$p-$n$h content of the collective states $| J k \rangle$
by calculating the amplitude $|\langle J k | J \text{$n$p-$n$h} \rangle|^2$.
Note that our procedure is unambiguous and can be used to analyze any state
constructed from mean-field wave functions. It permits to quantify the 
differences between simple mean-field configurations and the fully projected 
configuration mixing wave functions. The way $n$p-$n$h excitations are 
constructed is not the same as in shell model calculations and the 
single-particle wave functions do not have the same analytical form 
(the self-consistent wave functions discretized on a mesh have a much
better asymptotic behavior than oscillator wave functions). Nevertheless, 
our procedure will give us some insights on the relation between our 
wave functions and shell model ones.

The properties of the HF and projected HF states are given in 
table~\ref{tab:hf} where we also indicated the weight of the projected 
HF states in each of the four lowest collective $0^+$ states.

\begin{figure}[t!]
\centerline{\epsfig{file=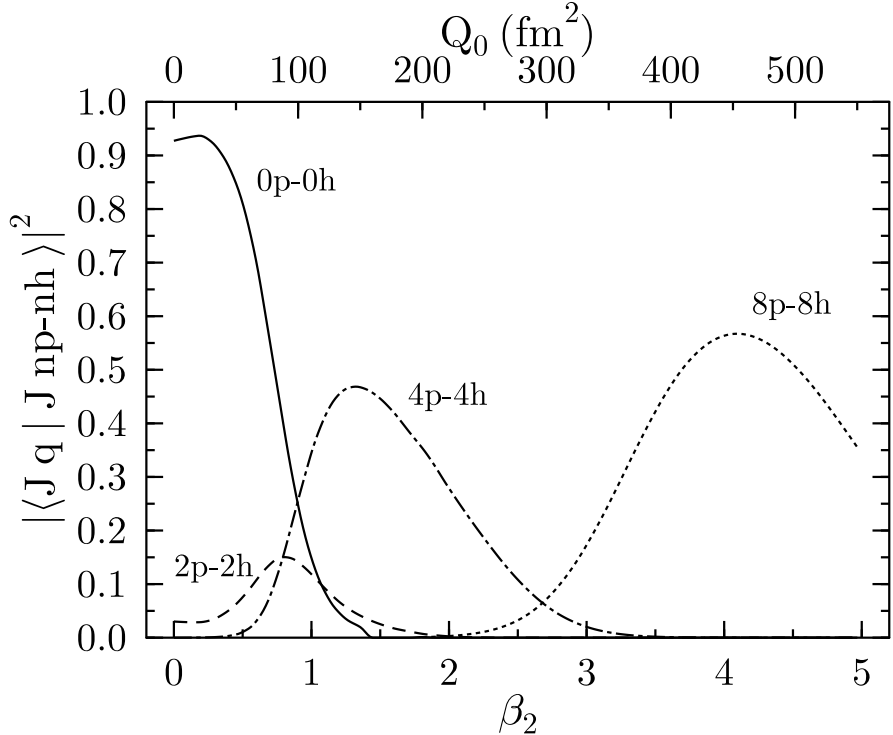,width=12cm}}
\caption{\label{fig:defhf}
Weight $|\langle J q | J \text{$n$p-$n$h} \rangle|^2$ of the \mbox{$J=0$}
projected spherical 0p-0h and deformed 2p-2h, 4p-4h, and 8p-8h HF states 
$| J \text{$n$p-$n$h} \rangle$ in the \mbox{$J=0$} projected paired mean-field 
states $| J q \rangle$. The weight of the two different 2p-2h states
is very similar, therefore just one is drawn.
}
\end{figure}

From figure \ref{fig:defhf}, one clearly sees that there is a correlation 
between the deformation of the projected mean-field state and its overlap 
with the $n$p-$n$h configurations. One sees also from table \ref{tab:hf}
that the deformation of the $n$p-$n$h HF states increases rapidly with the
number $n$ of excitations.

\begin{table}[t!]
\centerline{\begin{tabular}{lcccccccc} 
\noalign{\smallskip}\hline\noalign{\smallskip} 
state     & $E_{\text{HF}}$ 
          & $E_{J \text{HF}}$ 
          & $Q_0$
          & $\beta_2$
          & \multicolumn{4}{c}{$|\langle J k | J \text{$n$p-$n$h}\rangle|^2$} \\
\noalign{\smallskip}\cline{6-9}\noalign{\smallskip} 
          & (MeV) 
          & (MeV) 
          & (fm$^2$)
          &
          & $0^+_1$ & $0^+_2$ & $0^+_3$ & $0^+_4$ \\ 
\noalign{\smallskip}\hline\noalign{\smallskip}
0p-0h     &  0.0  &  0.0  &   0 & 0.00 & 0.737 & 0.195 & 0.006 & 0.0001 \\
2p-2h (p) & 14.92 & 11.96 &  41 & 0.37 & 0.094 & 0.098 & 0.018 & 0.003  \\
2p-2h (n) & 15.20 & 12.09 &  41 & 0.37 & 0.099 & 0.098 & 0.019 & 0.002  \\
4p-4h     & 15.95 & 11.26 & 119 & 1.08 & 0.027 & 0.212 & 0.193 & 0.003  \\
8p-8h     & 38.03 & 31.48 & 466 & 4.21 & 0.000 & 0.000 & 0.000 & 0.034  \\
\noalign{\smallskip}\hline\noalign{\smallskip}
\end{tabular}}
\caption{\label{tab:hf}
Excitation energy of the HF ($E_{\text{HF}}$) and \mbox{$J=0$} 
projected HF states ($E_{J \text{HF}}$) with respect to the (0p-0h) HF 
ground state. The mass quadrupole moment $Q_0$ and the deformation $\beta_2$
of the self-consistent 0p-0h, 2p-2h, 4p-4h, and 8p-8h HF states
are also given. The last four columns give the weights 
$|\langle J k | J \text{$n$p-$n$h} \rangle|^2$ of their \mbox{$J=0$} 
component in the four lowest $0^+$ states $| J k \rangle$
obtained from the configuration mixing.
}
\end{table}

With about 75 $\%$ weight of the 0p-0h HF state, 
see table \ref{tab:hf}, the breaking of the closed 
shell character of the $0^+_1$ ground-state wave function is not as 
large as in shell model calculations, but still sizable.
Note that a part of the admixture of higher shells obtained
in shell-model calculations is due to the fact that 
the ``real'' single-particle potential is not an harmonic oscillator 
as assumed in a shell model basis. Therefore 
it is not surprising that we find a smaller admixture of higher 
shells as we start with single-particle wave functions with a more 
realistic asymptotic behavior.

For the 4p-4h configuration, where a pair of protons and neutrons
is excited from the $1p_{1/2}$ to the $2d_{5/2}$ levels, a minimum 
in the HF energy is 
obtained for a mass quadrupole moment of \mbox{$Q_0 = 119$} fm$^2$, or 
\mbox{$\beta_2 = 1.08$}. This 4p-4h configuration corresponds to the
deformed \mbox{$N=Z=8$} shell closure visible at about the same deformation
in figure \ref{fig:s}. The excitation energy of this deformed
HF state (with respect to the spherical 
0p-0h HF state), is equal to 15.95 MeV, far above the 
experimental value of 6.03 MeV for the first excited $0^+$ state. 
The restoration of rotational symmetry by projection 
on angular momentum \mbox{$J=0$} decreases the calculated excitation 
energy to 11.25 MeV, still too high compared to experiment. 
To complete the analysis, we have also constructed 2p-2h and 8p-8h 
configurations. The only two 2p-2h states that we have considered are 
obtained by promoting either a pair of protons or a pair of neutrons 
from the $p_{1/2}$ shell to the $d_{5/2}$ shell. As we include 
the Coulomb interaction self-consistently, these two states have 
slightly different energies. The 8p-8h configuration
corresponds to the occupation of the $sd$ and $pf$ states which are 
below the Fermi level at very large deformations (see figure \ref{fig:s}).

%
%
%
%

The 0p-0h HF state is the dominant component of 
the collective ground state. If one 
adds the weights of the 0p-0h, 2p-2h and 4p-4h states, one obtains a
nearly complete description of the ground state, the missing components
being probably 2p-2h states not included in our analysis. One should,
however, take into account that the self-consistent $n$p-$n$h states
are not orthogonal and have small, but non-vanishing overlaps.
With less than 3$\%$, the contribution from the pure HF 4p-4h
configuration to the ground state is small and lower 
than the shell-model values (between 4 and 12$\%$ depending on 
the interactions). 

The excited $0^+_2$ state has a much more complicate structure,
without any dominant configuration. As can be seen in table 
\ref{tab:hf}, the weights of the 0p-0h, 2p-2h and 4p-4h 
configurations are quite close. These four components, however, represent 
only about 60$\%$ of the full state. Looking to figure \ref{fig:wf0},
one can see that the spreading of this state extends up to
deformations of the order of 200 fm$^2$, where the overlap
between the $n$p-$n$h HF states and the projected 
BCS states are small (see figure \ref{fig:defhf}). Many other
$n$p-$n$h configurations are therefore necessary to obtain a full
description of this $0^+_2$ state.

None of the collective states is dominated by the
deformed \mbox{$N=Z=8$} shell closure. The excited $0^+_3$ 
state has also a small 4p-4h component  and is constructed on
mean-field states with larger deformation than the 4p-4h HF state.

As can be seen from table \ref{tab:hf}, the 8p-8h state plays no role 
for the low-lying $0^+$ states. The first state in which it has a large weight 
of 0.48 is the $0^+_6$ state. The density distribution of the 8p-8h state is 
close to that of a four-$\alpha$-chain configuration. Its 38.0 MeV excitation 
energy is lowered to 32.2 MeV by projection on angular momentum and to 32.0 MeV 
by the configuration mixing. It is thus very large. This is, however, not 
inconsistent with the fact that there is up to now no convincing evidence 
for the presence of such a state in the low-energy spectrum of $^{16}$O. 
%
%
\section{Summary and Conclusions}
This study of the doubly-magic nucleus $^{16}$O has demonstrated
the descriptive and predictive power of methods based on a 
self-consistent mean-field approach including correlations.
In particular, the gain in energy of the pure 4p-4h HF
state, that is brought first by projection on angular
momentum, and then by mixing on the axial quadrupole moment,
is impressive. This procedure does not only permit to obtain 
the first excited $0^+$ state at the right energy, but also 
other excitations and transition matrix elements which are in 
reasonable agreement with available data. Somewhat surprisingly,
the 4p-4h HF state is evenly spread over the first and second excited
$0^+$ states, leaving none of them associated with a simple 4p-4h 
configuration.

Our results are not very sensitive to the parametrizations of the 
mean-field and pairing interactions that are used. The Skyrme 
interaction SLy4 used here was adjusted at the mean-field level
to binding energies, charge radii, and nuclear matter properties 
of a few nuclei, among them $^{16}$O. The correlations taken 
into account in our study increase the binding energy by 2.3 MeV 
compared to the spherical HF state, 
which leads now to a slight over-estimation of the $^{16}$O 
ground state-energy. Similarly, the proton rms radius is slightly
increased. Although it is gratifying to see that the change 
in total energy is below 2$\%$ while the change in radii stays 
below 3$\%$, it is clear that a readjustment of the effective 
interaction will have to be done when a larger experience will have 
been obtained on the effect of correlations.

We have included the most important symmetry restorations
to permit to compare our results directly with experimental
data in the laboratory frame of reference. Our calculations demonstrate
that this is a key to a successful quantitative description of the 
low-energy states when starting from a mean-field approach. Some 
ingredients, however, are still missing in our model. 
There are still symmetries broken by our approach.
Our wave functions break translational and Galilean invariances.
The implicit assumption is made that these broken symmetries bring similar
errors on all collective wave functions. To check this hypothesis in
our model is unfortunately still beyond numerical possibilities.
Projection on isospin seems to be the next natural step to enlarge 
the predictive power of the method even further. We have also not
considered proton-neutron pairing correlations, although $^{16}$O 
is an \mbox{$N=Z$} nucleus. Developments in the understanding of
these correlations and their modeling are still necessary. Concerning 
shape degrees of freedom, reflection-asymmetric configurations 
like $^{12}$C+$^4$He still have to be incorporated to describe several 
states of astrophysical interest. We have also restricted ourselves so far to 
axially symmetric shapes. Including triaxiality might alter the results
for high-lying excitations which spread differently into the 
$\beta$-$\gamma$ plane.
%
%
\subsection*{Acknowledgments}
This research was supported in part by the PAI-P3-043 of the Belgian
Office for Scientific Policy. We thank G.~Bertsch and H.~Flocard
for fruitful and inspiring discussions. M.~B.\ acknowledges support 
through a European Community Marie Curie Fellowship. P.-H.~H.\ thanks 
the Institute for Nuclear Theory at the University of Washington 
for its hospitality during the completion of part of this work.
%
%

\end{document}